\documentclass{webofc}
\usepackage[varg]{txfonts}   
\DeclareTextSymbol{\degreea}{T1}{6}
\newcommand{\degree}[0]{\degreea\,}
\usepackage[dvipsnames]{xcolor}
\usepackage[normalem]{ulem}

\begin{document}
\title{ Lattice-based Equation of State with 3D Ising  Critical point}
%
%

\author{\firstname{Micheal} \lastname{Kahangirwe}\inst{1}\fnsep\thanks{\email{mkahangi@central.uh.edu}} \and
        \firstname{Steffen} \lastname{A. Bass}\inst{2} \and
        \firstname{Johannes } \lastname{Jahan}\inst{1} \and
        \firstname{Pierre } \lastname{Moreau}\inst{2} \and
        \firstname{Paolo } \lastname{Parotto}\inst{3,4}
        \and
        \firstname{Claudia } \lastname{Ratti}\inst{1}
        \and
        \firstname{Olga } \lastname{Soloveva}\inst{5,6} 
        \and
        \firstname{Misha } \lastname{Stephanov}\inst{7,8,9} \and    
        \firstname{Elena } \lastname{Bratkovskaya}\inst{5,6,10}     
}

\institute{ Department of Physics, University of Houston, Houston, TX 77204, USA
\and
Department of Physics, Duke University, Durham, NC 27708, USA
\and
  Pennsylvania State University, Department of Physics, State College, PA 16801, USA
  \and
  Dipartimento di Fisica, Universit\`a di Torino and INFN Torino, Via P. Giuria 1, I-10125 Torino, Italy
\and
 Helmholtz Research Academy Hesse for FAIR (HFHF), 
 GSI Helmholtz Center for Heavy Ion Physics, Campus Frankfurt, 60438 Frankfurt, Germany
 \and
 Institut für Theoretische Physik, Johann Wolfgang Goethe-Universität, Max-von-Laue-Str. 1, D-60438 Frankfurt am Main, Germany  
\and
Kadanoff  Center  for  Theoretical  Physics,  University  of  Chicago,  Chicago,  Illinois  60637,  USA 
\and
  Physics Department, University of Illinois at Chicago, Chicago, IL 60607, USA 
\and
Laboratory for Quantum Theory at the Extremes, University of Illinois, Chicago, Illinois 60607,  USA
 \and
 GSI Helmholtzzentrum für Schwerionenforschung GmbH,Planckstrasse 1, D-64291 Darmstadt, Germany
}

\abstract{%

The BEST Collaboration equation of state combining lattice data with the 3D Ising critical point encounters limitations due to the truncated Taylor expansion up to $\frac{\mu_B}{T} \sim 2.5$. This truncation
consequently restricts its applicability at high densities.
Through a resummation scheme, the lattice results have been extended to $\frac{\mu_B}{T} = 3.5$.
In this article, we amalgamate these ideas with the 3D-Ising model, yielding a family of equations of state valid up to $\mu_B=700 \text{MeV}$ with the correct critical behavior. Our equations of state feature tunable parameters, providing a stable and causal framework-a crucial tool for hydrodynamics simulations.
}
\maketitle
\vspace{-1cm}
\section{Introduction}
\label{intro}
The primary objective of the Beam Energy Scan experiment at the Relativistic Heavy-Ion Collider (RHIC) is to map the Quantum Chromodynamics (QCD) phase diagram \cite{Aggarwal:2010cw,Bzdak:2019pkr}. Currently, researchers are in the process of analyzing the experimental data \cite{sweger2023recent}. On the theory side, at vanishing chemical potential, high-precision lattice simulations indicate that the transition from a hadron gas to a fluid of strongly interacting quarks and gluons is a smooth crossover \cite{aoki2006order, bazavov2012chiral}. Effective QCD-based models suggest that this crossover eventually transforms into a line of first-order transitions, featuring a critical point at finite density, the precise location of which is still unknown. Unfortunately, lattice simulations at finite baryon density are challenged by the sign problem.

Extrapolation methods that utilize lattice results at zero chemical potential are thus still needed for the equation of state at finite density. Taylor expansion is widely used and is effective for low-density physics, approximately $\mu_B/T < 3$; \cite{allton2002qcd, kaczmarek2011phase, bollweg2022taylor}. However, its applicability is limited, making it difficult to model high-density critical phenomena.

The BEST Collaboration constructed a parametric family of possible equations of state which, at zero $\mu_B$, agree with the lattice Taylor expansion data, while featuring a critical point at chosen $T_C$ and $\mu_{BC}~$ \cite{parotto2020qcd}. However, the construction appears to be limited to $\mu_B\lesssim 450$ MeV due to unphysical features ("wiggles") induced by extrapolation of Taylor expansion to higher $\mu_B$. On the other hand, the range of $\mu_B> 450$ MeV is needed to describe hydrodynamics in the regime relevant for RHIC Beam Energy Scan.

The Wuppertal-Budapest collaboration developed a novel re-summation expansion scheme \cite{Borsanyi:2021sxv} that exhibits smooth behavior at high density, producing no visible artifacts up to $\mu_B/T \sim 3.5$, by introducing a chemical potential-dependent transition temperature. This scheme simply relates the baryon density $\chi_1^B(T, \mu_B)\equiv n_B(T,\mu_B)$ with the baryon susceptibility $\chi_2^B(T, \mu_B=0)$ at zero chemical potential:
\begin{equation}
T\frac{\chi_1^B(T,\mu_B)}{\mu_B} = \chi_2^B(T',0) \;,
~~~\text{where}~~~ 
\chi_n^B(T) = \frac{\partial^n P/T^4 }{\partial (\mu_B/T)^n} \;,
\label{Eq:AltExS}
\end{equation}
via an effective temperature
\begin{equation}
\label{eq:T'-definition}
T'(T,\mu_B) = T\left(1+\kappa_2^{BB}(T)\left(\frac{\mu_B}{T}\right)^2 + \kappa_4^{BB}(T)\left(\frac{\mu_B}{T}\right)^4 + \ldots\right) \;.
\end{equation}

This scheme has inherent advantages, because $\kappa_2^{BB}(T)$ does not introduce wiggles since it is constant near the transition, and $\kappa_4^{BB}$ is consistent with zero. The remaining task is to introduce a critical point in this equation of state.

\subsection{Methodology }\label{sec:methodology}
\subsubsection{Mapping Ising to QCD}\label{subsec:mapping}

\begin{figure*}[!htbp]
\centering
 \includegraphics[scale=0.2]{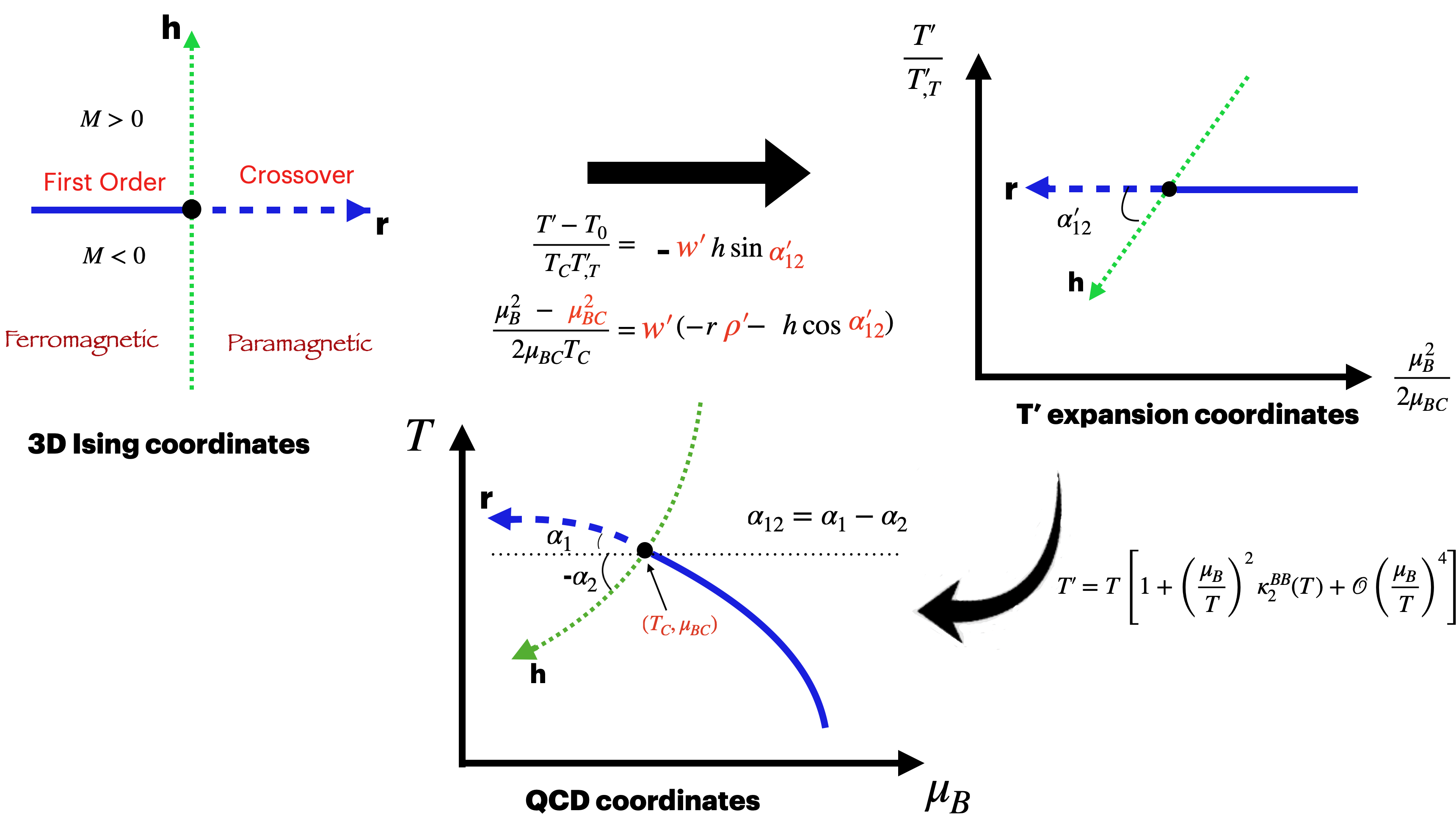}
\caption{The top left plot represents the 3D Ising model with a critical point located at $(r=0, h=0)$. The top right plot displays the alternative $T$-expansion scheme coordinates, with a critical point at $(T' = T_0, \mu_B = \mu_{BC})$. Finally, the bottom plot corresponds to the QCD coordinates, featuring a critical point located at $(T_C, \mu_{BC})$. The tunable (yet unknown) parameters are in red.}
\label{fig:mapping}
\end{figure*}

Near the critical point, the divergence of the correlation length results in a system whose behavior is determined largely by (global) symmetries, and not by microscopic degrees of freedom. Systems with shared symmetries exhibit similar critical behavior. 
This fact has been used in Ref. \cite{parotto2020qcd} to map the universal critical behavior from the 3D Ising model to the vicinity of the QCD critical point. To combine the approach of Ref. \cite{parotto2020qcd} with the Taylor resummation approach of Ref. \cite{Borsanyi:2021sxv}, in this work we employ a two-step mapping with the intermediate step using $(T', \mu_B)$ coordinates, where the line of first-order phase transition remains constant at $T'=T_0$.
Upon substitution of $T'_{lattice}$ for $T'$ in the subsequent step, the critical line follows the curvature in the QCD coordinates, as illustrated in Fig. \ref{fig:mapping}.

The mapping depicted in Fig. \ref{fig:mapping} involves free parameters: $w'$, $\rho'$, and $\alpha_{12}^\prime$ which can easily be related to the BEST parameters $w$, $\rho$, and $\alpha_{1}$ and $\alpha_{2}$ in \cite{parotto2020qcd}, as detailed in \cite{Kahangirwe:2024cny}, Being inherently even in $\mu_B$, this mapping ensures baryon-antibaryon symmetry. In Section \ref{sec:mergingIsing}, we explain how to merge the critical behavior with the lattice data.

\subsubsection{Embedding Ising Critical point into lattice equation of state}\label{sec:mergingIsing}
It is crucial to interpret Eq. \eqref{Eq:AltExS} as a definition of $T'$. The physics at finite density should be incorporated into $T'$, including the singularity at the critical point. To ensure smooth behavior, we initially merge the lattice data $\kappa^{BB}_{2,\text{lattice}}$ and $\chi^{BB}_{2,\text{lattice}}$ with HRG data (based on the PDG 2021+ particles list \cite{SanMartin:2023zhv}), and parameterize in \cite{Kahangirwe:2024cny} to cover the temperature range relevant for hydrodynamics. Using Eq. \eqref{Eq:AltExS}, we construct $n_B^{full}$ as follows:

\begin{equation}
     \frac{n_B(T,\mu_B)}{T^3}=\frac{\chi_1^B(T,\mu_B)}{T^3} = \left(\frac{\mu_B}{T}\right)\chi^B_{2,\text{lattice}}(T'_{\text{full}}(T,\mu_B),0) \label{baryonFull}
\end{equation}
\begin{eqnarray} 
    T'_{\text{full}}(T,\mu_B) &= \underbrace{T'_{\text{lattice}}(T,\mu_B)}_{\text{lowest orders in $(\mu_B/T)$ }} ~ + ~
\underbrace{T'_{\text{crit}}(T,\mu_B) - \text{Taylor}[T'_{\text{crit}}(T,\mu_B)]}_{\text{higher order in $(\mu_B/T)$ }}\label{Tfull}.
\end{eqnarray}

$T'_{\text{lattice}}$ is kept to $\mathcal{O}((\mu_B/T)^2)$, and the critical part contributes to higher orders. $T'_{\text{crit}}$ is:
\begin{align}
    T'_{\text{crit}}(T,\mu_B) \approx &  \left(\frac{\partial \chi^B_{2,\text{lattice}}(T,0)}{\partial T}\Big|_{T_0}\right)^{-1} \frac{n^{\text{crit}}_B(T,\mu_B)}{(\mu_B/T)} 
    \label{Tcrit}
\end{align}
where $n_B^{\rm crit}$ is the singular contribution to $n_B$ obtained by mapping the 3D Ising model to QCD coordinates.
By substituting Eq. \eqref{Tcrit} into Eq. \eqref{Tfull}, we reconstruct the full baryon density using Eq. \eqref{baryonFull}.

\subsection{Results }\label{sec:results}

\begin{figure}[!htbp]
\centering
\includegraphics[width=6.459cm,clip]{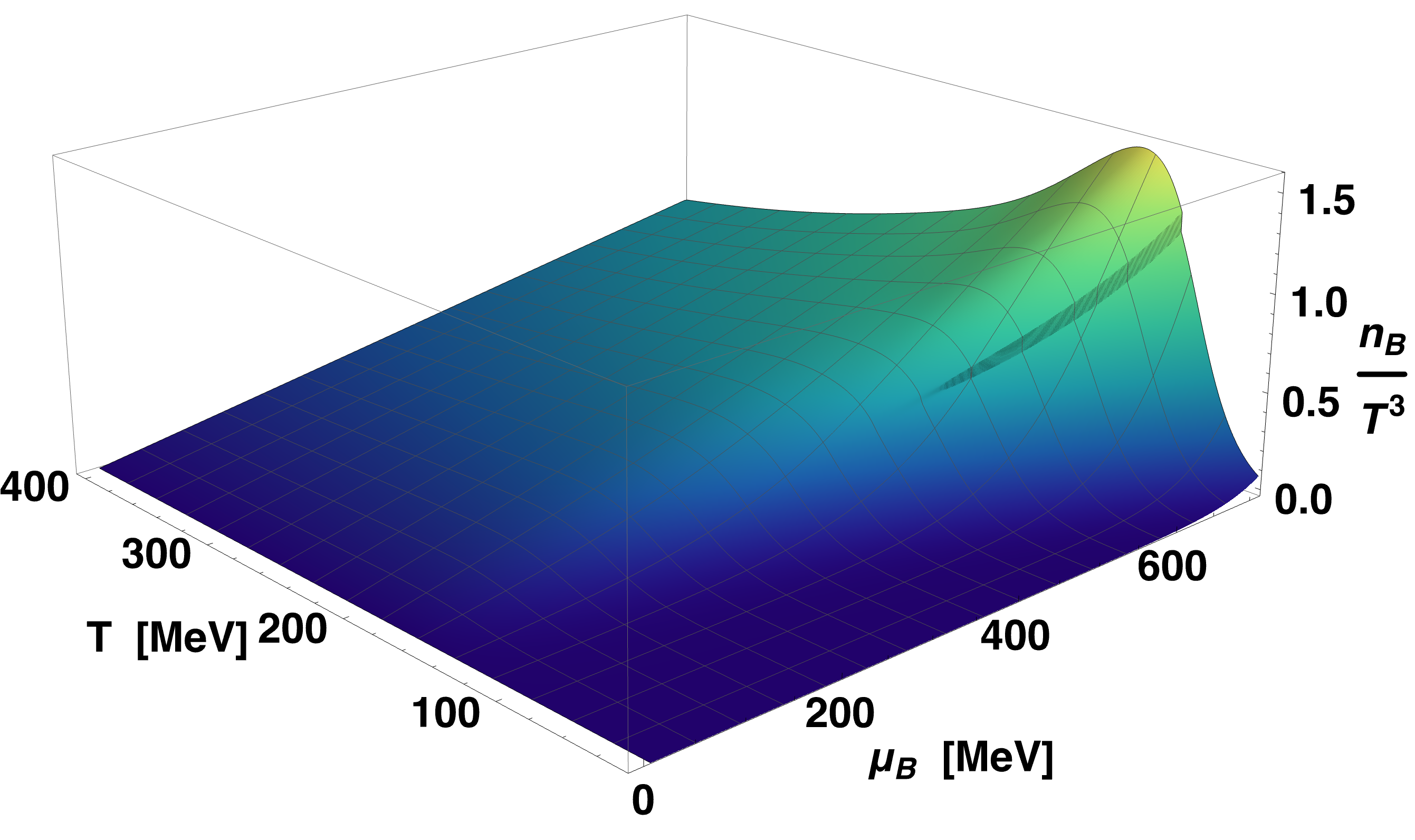}
\includegraphics[width=6.cm,clip]{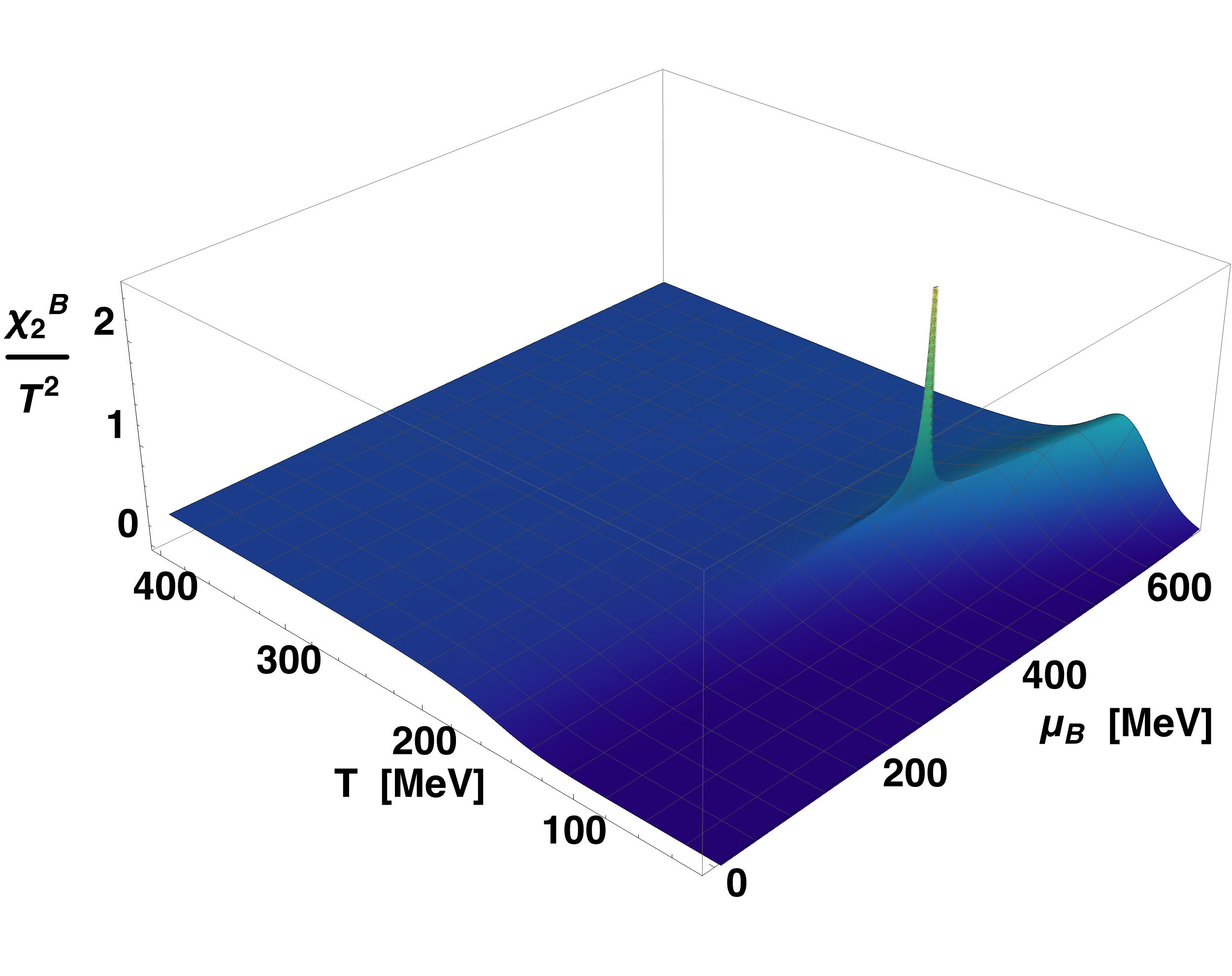}
\caption{Left panel: baryon density, with a discontinuity marking a first-order transition. Right panel: $2^\text{nd}$ order baryon susceptibility with a divergence indicating the critical point. Both have been obtained using parameters $\mu_{BC}=500 \text{MeV}$, $T_C=116.49 \text{MeV}$, $w =15.0$, $\rho=0.3$, $\alpha_1 = 11.19$\degree and $\alpha_{12} =\alpha_1 $. }
\label{fig:thermodynamics}       
\end{figure}

In this contribution we present an enhancement in QCD thermodynamics with respect to the BEST collaboration result \cite{parotto2020qcd}, as illustrated in Fig. \ref{fig:thermodynamics}.  This improvement extends the analysis to a higher chemical potential, reaching up to $\mu_B = 700 ~\text{MeV}$. The transition line in the figure exhibits the expected curvature in QCD, thereby reducing the number of free parameters. Specifically, we fix $\alpha_1$ to be the slope of the curve in Fig. \ref{fig:mapping}. Furthermore, our parametric equations of state remain valid for smaller mapping angles $\alpha_{12}$, corresponding to physical (small) quark masses, according to \cite{Pradeep:2019ccv}.

\subsection{Acknowledgements}
This work was supported in part by the National Science Foundation (NSF) under grants No. PHY-2208724 and PHY-2116686, and within the framework of the MUSES collaboration, under grant number OAC-2103680, and by the U.S. Department of Energy, Office of Science, Office of Nuclear Physics, grants No. DEFG0201ER41195, DE-SC0022023, DE-FG02-05ER41367 and Deutsche Forschungsgemeinschaft through the grant CRC-TR 211, Project N 315477589 - TRR 211

\bibliography{reference}

\end{document}